\documentclass[%
 reprint,
superscriptaddress,
 amsmath,amssymb,
prl,
]{revtex4-2}

\usepackage{graphicx}
\usepackage{dcolumn}
\usepackage{bm}
\usepackage{xcolor}
\usepackage[english]{babel}
\usepackage{ulem}

\begin{document}


\title{Control of the phonon band gap with isotopes in hexagonal boron nitride}

\author{Paul Zeiger}
\email{paul.zeiger@physics.uu.se}
\affiliation{Department of Physics and Astronomy, Uppsala University, L\"{a}gerhyddsv\"{a}gen 1, 751 20 Uppsala, Sweden}

\author{Jordan Hachtel}
\affiliation{Center for Nanophase Materials Sciences, Oak Ridge National Laboratory, Oak Ridge, TN 37830, USA}

\author{Dominik Legut}
\affiliation{IT4Innovations, V\v{S}B - Technical University of Ostrava, 17. listopadu 2172/15, 70800 Ostrava-Poruba, Czech Republic}

\author{Eli Janzen}
\affiliation{Tim Taylor Department of Chemical Engineering, Kansas State University, Manhattan, Kansas 66506, USA}

\author{Juri Barthel}
\affiliation{Ernst Ruska-Centre (ER-C 2) Forschungszentrum Juelich GmbH, 52425 Juelich, Germany}

\author{James H. Edgar}
\affiliation{Tim Taylor Department of Chemical Engineering, Kansas State University, Manhattan, Kansas 66506, USA}

\author{Leslie J. Allen}
\affiliation{School of Physics, University of Melbourne, Parkville, VIC 3010, Australia}

\author{J\'{a}n Rusz}
\affiliation{Department of Physics and Astronomy, Uppsala University, L\"{a}gerhyddsv\"{a}gen 1, 751 20 Uppsala, Sweden}

\date{\today}

\begin{abstract}
The isotopic mass of constituent elements of materials has a well-known effect on the energy of vibrational modes. By means of monochromated scanning transmission electron microscopy we have experimentally studied the phonon bandstructure of hexagonal BN, where a phonon band gap appears between in-plane optical phonon modes and the lower energy part of the phonon spectrum. The size of the phonon band gap can be manipulated by the isotopic mass of the boron. While in $^{11}$BN the phonon band gap is about 7~meV wide, in $^{10}$BN the gap nearly closes, being an order of magnitude smaller (below 0.5~meV). This opens exciting options for manipulating terahertz wave propagation through isotopically structured devices having otherwise no interfaces between chemically distinct components.
\end{abstract}

\maketitle


Isotopic composition of materials has a well-known effect on the vibrational modes of materials. In a simple linear chain model -- a classical textbook example -- a band gap in vibrational modes opens when the chain consists of two alternating atoms with different masses \cite{kittel_introduction_2004}. The atoms can be either of different species, or of the same species but different isotopes. In real materials, by manipulation of isotopic composition, we can induce non-trivial changes of the phonon band structure \cite{cardona_isotope_2005}. In materials in which several stable isotopes co-exist, the resulting disorder provides an additional scattering mechanism, strongly influencing the thermal conductivity. In turn, working instead with isotopically pure materials can greatly enhance the thermal conductivity, as was seen in hexagonal boron nitride (h-BN) \cite{lindsay_enhanced_2011,Yuan_modulating_2019}. Via electron-phonon coupling, the changes of phonon modes due to different isotopic masses can transfer even to electronic properties, for example influencing an electronic band gap \cite{plekhanov_isotopic_1997,plekhanov_isotope_2003,vuong_isotope_2018}.

\begin{figure}[h!]
    \centering
    \includegraphics[width=\columnwidth]{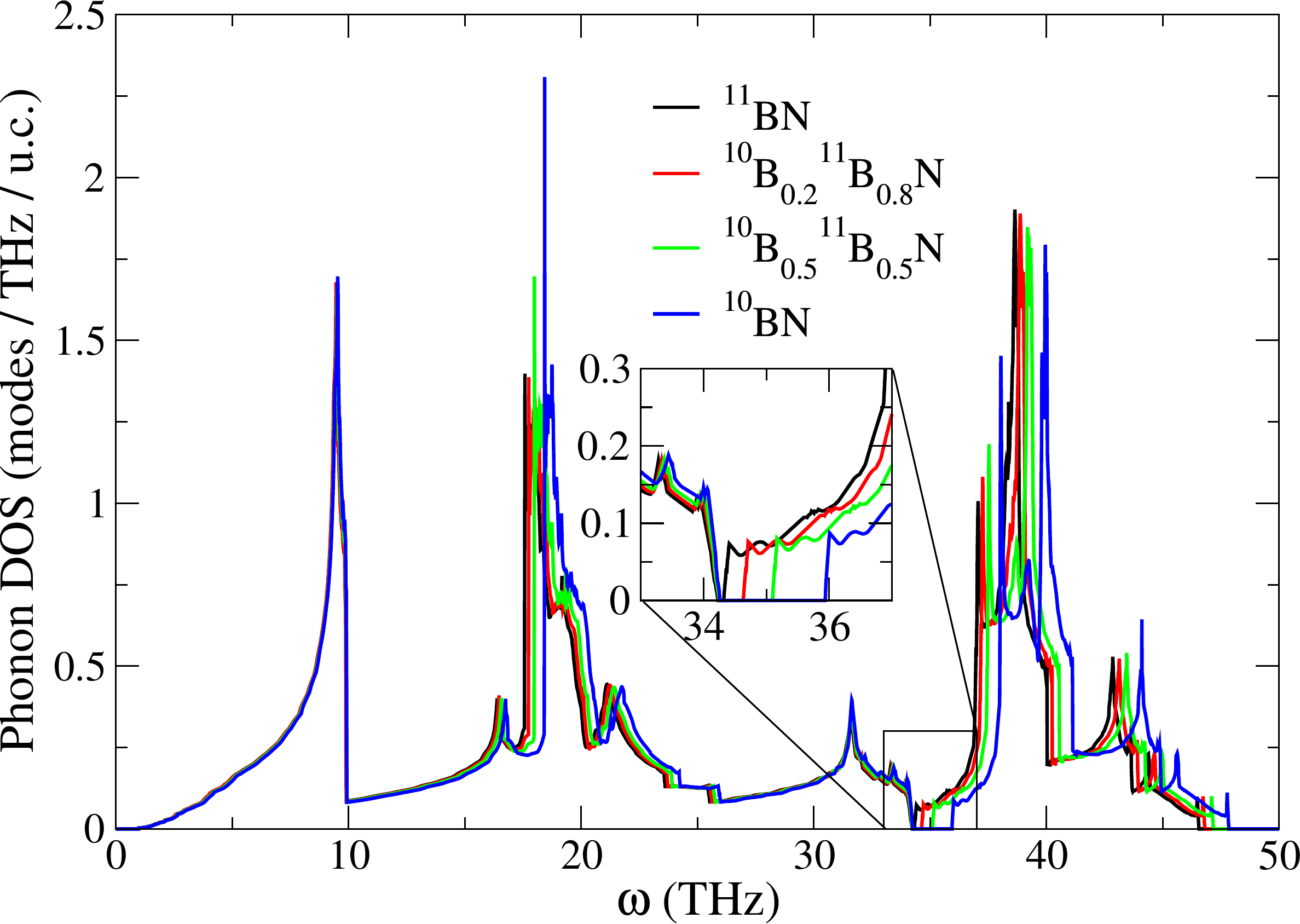}
    \includegraphics[width=\columnwidth]{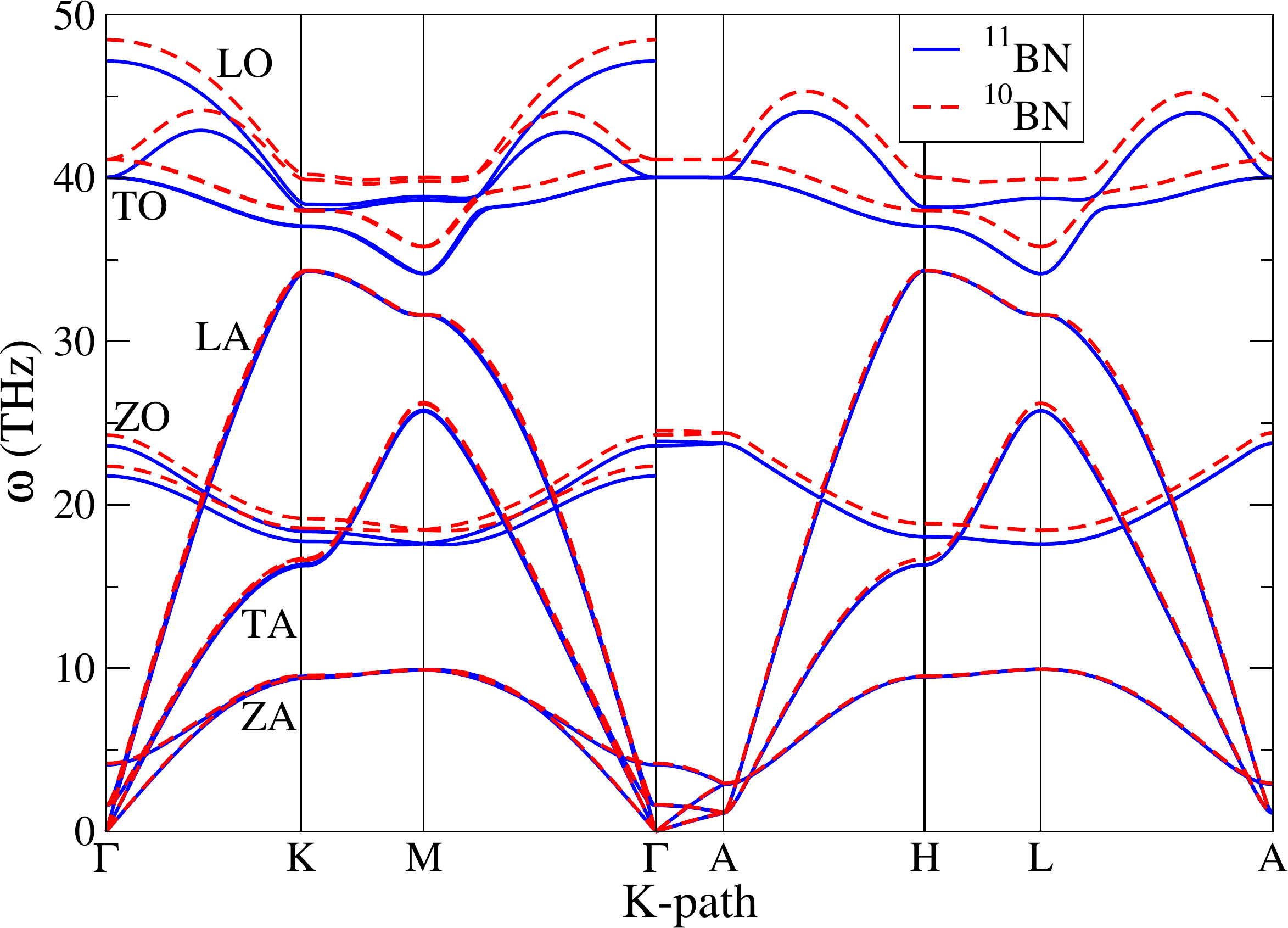}
    \includegraphics[width=3cm]{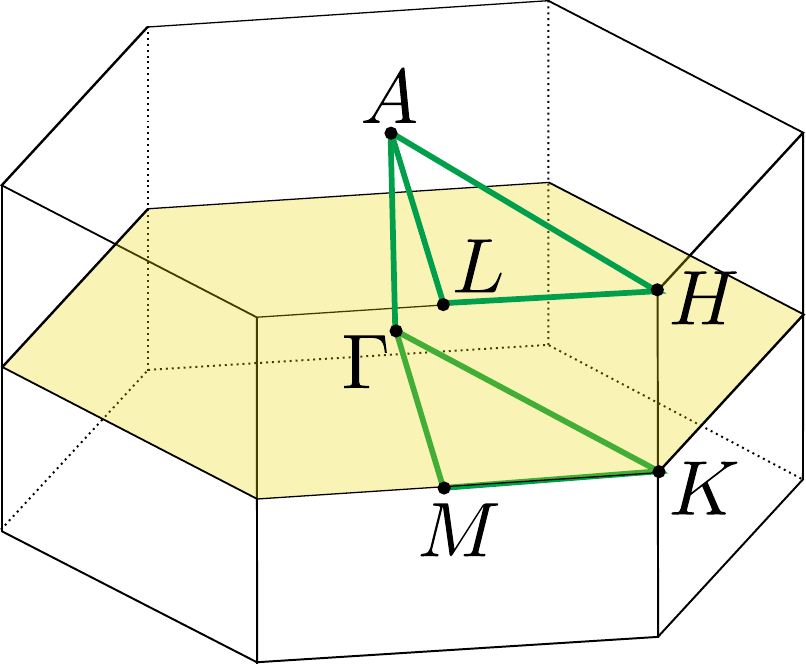}
    \includegraphics[width=4cm]{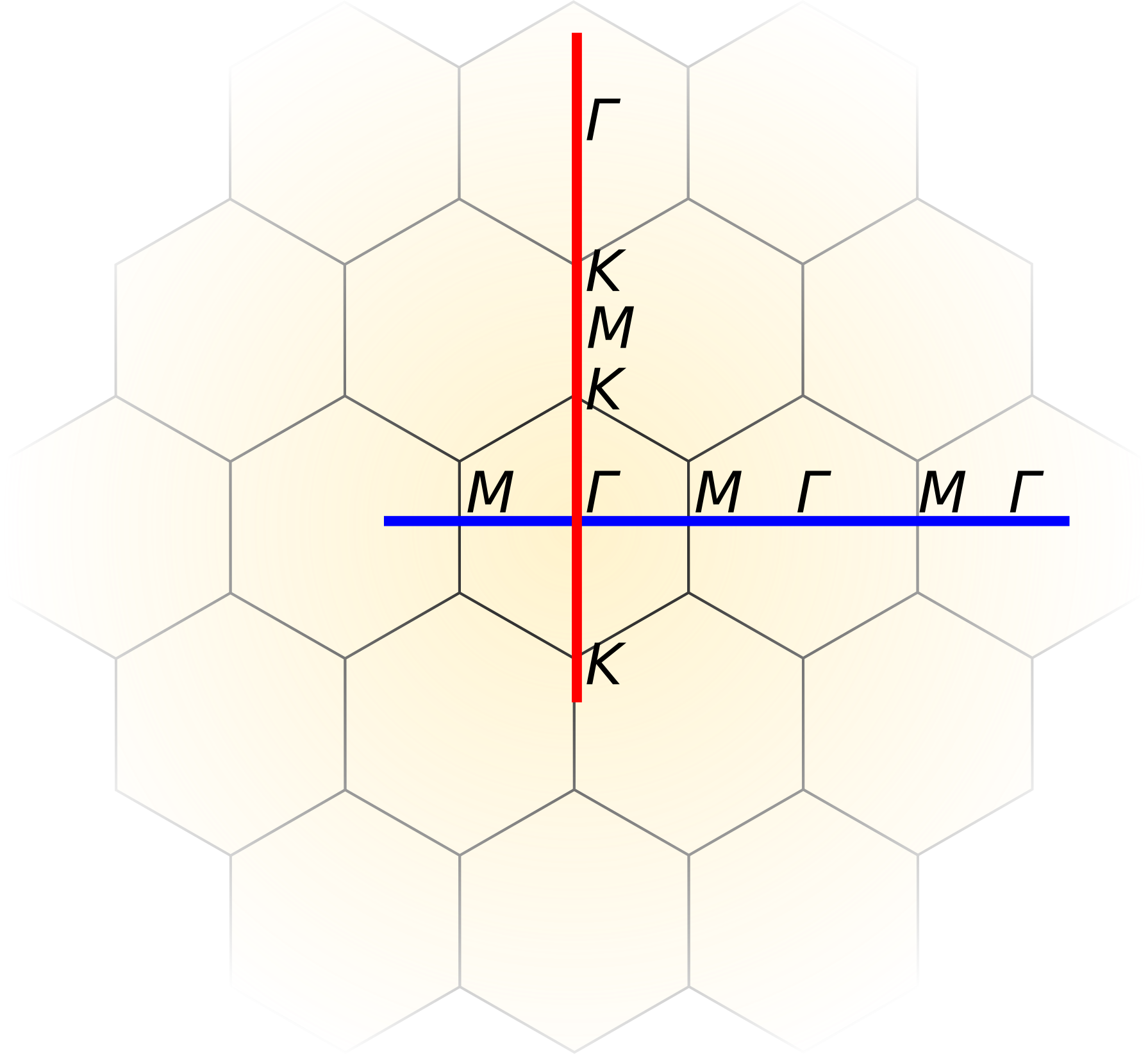}
    \caption{DFT calculations of the phonon DOS of h-BN with various B isotope proportions. The phonon bandgap between 34 and 36~THz (see inset) increases with increasing $^{10}$B concentration. Phonon band structures of $^{11}$BN(blue solid line) and $^{10}$BN (red dashed line) are plotted in the middle panel, including the LO--TO splitting (note the discontinuities at $\Gamma$ at the beginning of the $\Gamma$--$A$ segment) and annotation of individual bands. Bottom panel shows a plot of the Brillouin zone of h-BN with highlighted special points. In addition, we indicate lines along which AREELS data were calculated. The center of the diffraction pattern is at the crossing of the blue and red lines.}
    \label{fig:PDOS}
\end{figure}

Recent advances in instrumentation enabled the use of scanning transmission electron microscopes (STEMs) to study vibrational modes of materials \cite{krivanek_vibrational_2014, 3DPhonons, STOGB}. This comes together with the advantages of a STEM offering a spatial resolution superior to other experimental methods, reaching the atomic scale \cite{STOCTO,yan_stacking_fault_2021,hage_single-atom_2020,venkatraman_vibrational_2019}. Energy resolution of these monochromated STEMs is also sufficient for isotopic labeling, as was demonstrated both theoretically \cite{konecna_theory_2021} and experimentally \cite{IsotopicGraphene,hachtel_identification_2019,H20D20}.



In this Letter we show that by changing the isotopic composition in h-BN it is possible to manipulate its phonon band gap. Isotopically pure cases of $^{11}$BN and $^{10}$BN manifest a phonon band gap of magnitude between 0.5~meV and 7~meV, separating in-plane optical phonon modes from a continuous band consisting of the out-of-plane and in-plane acoustic modes, with a possibility to control the gap size by choosing a suitable mixture of the two stable isotopes of boron.





We have performed density functional theory (DFT) calculations of the electronic structure to extract the phonon density of states for different isotopic compositions. The results were obtained using the projected augmented waves potentials with electronic valence configurations of $2s^2 2p^1$ and $2s^2 2p^3$ for B and N atoms 
as implemented in the VASP code \cite{VASP}. The electronic correlation  effects were treated by the generalized gradient method as parametrized by Perdew-Burke-Ernzerhof \cite{PBE}, the electronic exchange effects including the one stemming from the van der Waals interactions by the approach reported in Ref.~\cite{vdWaals}. The number of plane waves was limited by the energy cut-off of 700 eV, the k-point sampling was done by the $\Gamma$ centered grid of 9$\times$9$\times$9 $\mathbf{k}$-points, the total energy and Hellman-Feynman forces were converged to $10^{-8}$ eV and $10^{-4}$ eV/\AA, respectively. The lattice dynamic was calculated by the means of the direct method \cite{Parlinski} that utilizes the DFT calculated Hellmann-Feynman forces acting on all atoms within the supercell of $2 \times 2 \times 1$ cells (16 ions) with displacements of 0.01~\AA{} for symmetry non-equivalent atoms using the \textsc{Phonopy} code \cite{phonopy}. The dipole-dipole interaction (non-analytic term correction nearby the $\Gamma$ point), {\it i.e.}, the splitting of the longitudinal and acoustic optical modes was also taken into account \cite{Gonze-LOTO}.

The results are summarized in Fig.~\ref{fig:PDOS}. As we increase the concentration of the $^{10}$B isotope, the in-plane optical phonon modes (approximately above 35~THz) move up in energy. A similar effect, albeit less pronounced, also occurs for out-of-plane modes nearby a frequency of 20~THz (ZO modes). As the top of the frequency range of longitudinal acoustic (LA) modes (30--34~THz) stays almost unchanged for all isotopic compositions, the in-plane optical phonon modes detach and a gap in the phonon density of states (PDOS) grows with decreasing isotopic weight of boron. For $^{11}$BN there is a small ($<0.5$~meV) gap, which widens at the natural isotopic composition $^{10}$B$_{0.2}\phantom{}^{11}$B$_{0.8}$N. PDOS at the natural isotopic composition agrees with results in the literature \cite{kern_ab_initio_1999}. Increasing the $^{10}$B concentration further to $^{10}$B$_{0.5}\phantom{}^{11}$B$_{0.5}$N widens the gap to 4~meV, eventually reaching approximately 7~meV for isotopically pure $^{10}$BN.

Phonon dispersion relations in Fig.~\ref{fig:PDOS} demonstrate in a $\mathbf{q}$-resolved way that it is indeed the LO and TO bands which shift with changing isotopic composition, while the LA band remains practically unchanged. To observe this effect, measurements must access the $\mathbf{q}$-resolved phonon band structure. As was outlined above, STEMs can detect phonon modes with an energy resolution sufficient for distinguishing isotopes of light elements \cite{hachtel_identification_2019}. Moreover, in the angle-resolved electron energy loss spectroscopy (AREELS) mode, where a probe with a very small convergence angle is used, it allows to measure phonon band dispersions with high spatial resolution \cite{qi_measuring_2021,plotkin-swing_hybrid_2020,senga_position_2019}. Therefore we have turned our attention to an AREELS experiment.

\begin{figure*}[t!]
    \centering
    \includegraphics[width=14cm]{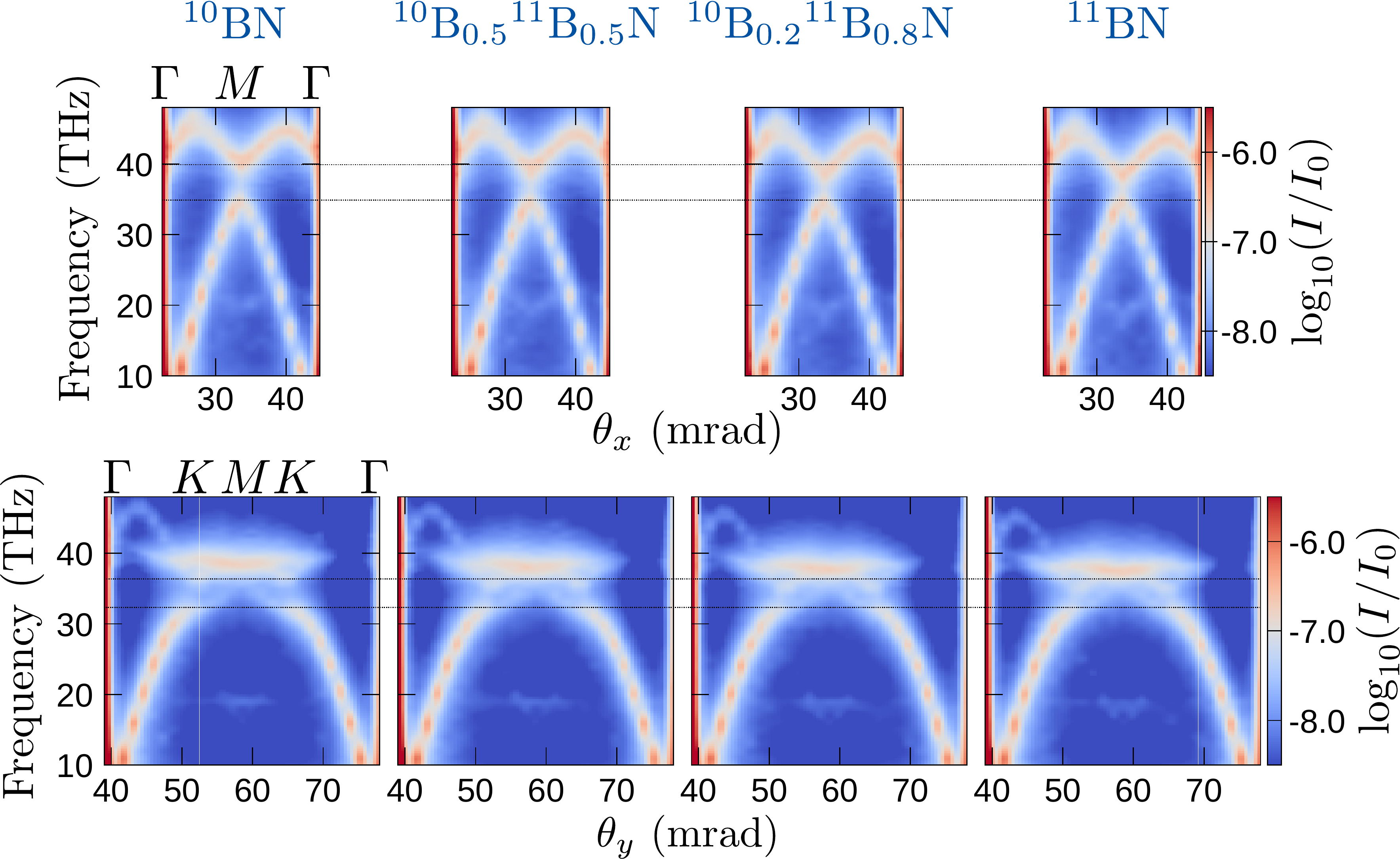}
    \caption{Calculated AREELS along the $\Gamma$--$M$--$\Gamma$ (top row) and $\Gamma$--$K$--$M$--$K$--$\Gamma$ (bottom row) paths in the Brillouin zone. The proportion of $^{11}$B isotope increases from left to right. Horizontal dotted lines serve as a guide for the eye, marking the top of the LA band at the $M$-point and the bottom of the $TO$ band at the $M$-point (top) or the $K$-point (bottom), respectively, in $^{10}$BN.}
    \label{fig:FRFPMS}
\end{figure*}

Before reporting our experimental results, we first present simulations of the AREELS datasets by means of frequency-resolved frozen phonon multislice (FRFPMS) method \cite{zeiger_efficient_2020,zeiger_frequency-resolved_2021}, which captures the dependence of the spectral shapes on phonon polarization vectors as well as the dynamical diffraction effects \cite{zeiger_theory_2023}. FRFPMS requires snapshots of the vibrating structure as an input. We have performed molecular dynamics simulations with the \textsc{lammps} code \cite{plimpton_fast_1995}. An orthogonal simulation box consisting of $16 \times 14 \times 50$ unit cells of h-BN was considered with dimensions $a=3.47$~nm, $b=3.51$~nm and $c=32.9$~nm, which were determined by energy minimization of the structure. In all MD simulations, the time step was $0.5$~fs. To maintain near-DFT level accuracy in the phonon dynamics, the interatomic potential was described by a machine-learned, so-called Gaussian approximation potential (GAP) \cite{thiemann_machine_2020}. This provides a much more accurate description of vibrational properties, compared to our previous works \cite{zeiger_efficient_2020,zeiger_frequency-resolved_2021,zeiger_simulations_2021}. A total of 49 frequencies were simulated. These include the range 10-48~THz in steps of 1~THz and, additionally, the half-THz frequencies between 35 and 45~THz to cover the in-plane optical phonon bands in more detail. The multislice calculations were performed using \textsc{DrProbe} \cite{barthel_dr_2018} using ionic potentials \cite{hage_contrast_2020} in a full three-dimensional mode, i.e., without employing the usual $z$-projection approximation. Multislice grid was set to $448 \times 448 \times 1000$ voxels and the calculations included Debye-Waller factors calculated by \textsc{lammps} at room temperature, as well as absorption factors \cite{weickenmeier_computation_1991}. An acceleration voltage of 60~kV was assumed and the incoming beam used in AREELS experiments (see below) was approximated by a plane wave propagating parallel to the $c$-axis of the h-BN unit cell.

Simulated AREELS for different isotopic compositions are summarized in Fig.~\ref{fig:FRFPMS}. Consistent with Fig.~\ref{fig:PDOS}, we see that the LO and TO bands are shifting towards higher energies as the $^{10}$B isotope concentration increases. Note that the energy shifts in the ZO modes are visually absent in our simulations. This is due to the low sensitivity of the EELS to the vibrations along $z$-direction, where polarization vectors are almost perpendicular to momentum transfers \cite{nicholls_theory_2019}. Energy shifts of the in-plane optical modes are clearly visible in the simulations, suggesting that the effect should be detectable in experiments.

Bulk samples of hBN were grown at atmospheric pressure from a molten iron-chromium solution using $^{10}$B and $^{11}$B isotopically enriched boron sources \cite{hBNPrep}. Thin flakes prepared by chemical exfoliation were then transferred onto lacey carbon TEM grids. Flakes covering a wide range of sizes were present. Samples for AREELS were chosen based off of large area flakes with minimum strain contrast in the Kikuchi diffraction pattern.

AREELS was acquired on a Nion aberration-corrected high-energy-resolution monochromated EELS-STEM (HERMES) equipped with a Nion Iris Spectrometer. A convergence angle of ~2-3 mrad was used for the experiments with a 3~mm $\times$ 0.5~mm slot aperture used for collection. A collection length (not angle since the slot aperture is effectively one-dimensional) is set to ~30 mrad to capture a full BZ effectively, which corresponds to a collection width of ~5 mrad. We also acquire the AREELS from the second BZ away from the optic axis which usually exhibits the clearest signal. Due to the difficulty of the experiments the AREELS datasets used for this manuscript came from separate experiments, one conducted at an accelerating voltage of 60 kV ($^{10}$BN) and the other at 30 kV ($^{11}$BN). Both experiments had an energy resolution of approximately 30 meV, as measured by the full-width at half-maximum (FWHM) of the quasi-elastic scattering zero peak directly below the M' point. The differing accelerating voltages may influence the scattering cross-sections of the phonons, but should not significantly change the measured phonon frequencies at any momentum. 

\begin{figure}[t!]
    \centering
    \includegraphics[width=8.6cm]{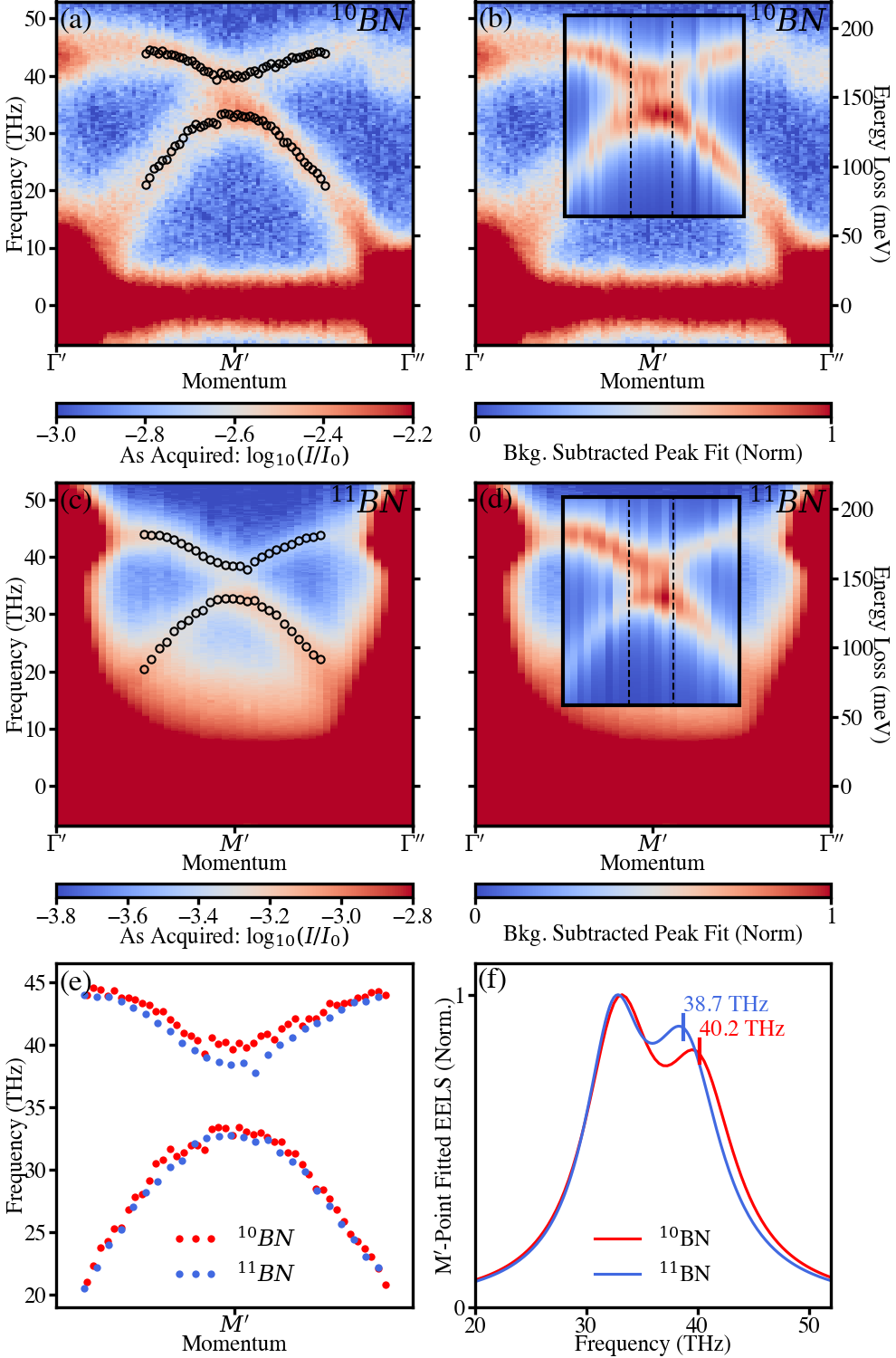}
    \caption{Measured phonon dispersions along the $\Gamma$--$M$--$\Gamma$ path for $^{10}$BN (a,b) and $^{11}$BN (c,d). The As-acquired EELS data (a,c) is fitted with four Lorentzians (two to capture the background of the ZLP, two to capture the band-dispersion), enabling direct access to the band dispersions without any background (b,d). The background-subtracted peak-fitted band-positions are overlaid in (e) demonstrating that the energies of the longitudinal acoustic band stays almost unchanged for both compositions, while the longitudinal optical band shifts to higher energies in isotopically pure $^{10}$BN. The channels closest to the $M^\prime$ point are averaged in (f), and show a shift of the optic band by 1.5 THz (5.9 meV).}
    \label{fig:experiment}
\end{figure}

Figure~\ref{fig:experiment} shows measured AREELS for $^{10}$BN and $^{11}$BN isotopes along the $\Gamma^\prime-M^\prime-\Gamma^{\prime\prime}$ line (see Fig.~\ref{fig:PDOS}). Empty and filled circles mark peak-fitted positions of the LA and LO bands, to which the measurements are most sensitive in our experimental geometry, thanks to a parallel orientation of momentum transfer and phonon polarization vectors of longitudinal modes \cite{nicholls_theory_2019,zeiger_frequency-resolved_2021}. By comparing the phonon band energies for the two isotopic compositions, we demonstrate the predicted isotopic shift of the LO mode to higher energies in a $\mathbf{q}$-resolved way. Such angle-resolved energy shifts provide an unprecedented detailed view of phonon band-structure modifications due to varying isotopic composition. We also note that at the 60~kV acceleration voltage used here and 2-3~mrad probe-forming convergence semi-angle, the electron beam size remains well below 2~nm. The experimentally observed energy shift between $^{10}$BN and $^{11}$BN is of the largest magnitude in the vicinity of the $M^\prime$-point and has a magnitude of approximately 1.5 THz/5.9 meV, which matches well with our predictions of a 1.6 THz/6.5 meV shift.


In conclusion, we have theoretically predicted and experimentally demonstrated a modification of phonon band structure as a function of the boron isotopic composition of hexagonal boron nitride. Angle-resolved electron energy loss spectroscopy has revealed the predited sub-10~meV shifts of phonon band energies in a momentum-resolved way. We have shown that the phonon band gap in hBN can be controlled within more than an order of magnitude by manipulating isotopic composition. This opens up the possibility of exciting applications for guiding terahertz waves within isotopically structured materials, where certain frequencies are transferred without resistance, while other can be selectively guided or blocked.

\begin{acknowledgments}
We acknowledge the support of Swedish Research Council, Olle Engkvist's foundation and Carl Trygger's Foundation. Part of the simulations was enabled by resources provided by the Swedish National Infrastructure for Computing (SNIC) at NSC Centre partially funded by the Swedish Research Council through grant agreement no.\ 2018-05973. DL acknowledges projects e-INFRA CZ (ID:90140) and INTER-EXCELLENCE II (LUASK22099) supported by the Ministry of Education, Youth and Sports of the Czech Republic and the project 23-07228S of the Czech Science Foundation. We acknowledge the resources given by the LUMI-C pilot phase, project number 465000027. Support for the hBN crystal growth came from the Office of Naval Research, award number N00014-22-2582. Momentum-resolved EELS was conducted at the Center for Nanophase Materials Sciences, which is a US Department of Energy Office of Science User Facility using instrumentation within ORNL’s Materials Characterization Core provided by UT-Battelle, LLC, under Contract No. DE-AC05- 00OR22725 with the DOE and sponsored by the Laboratory Directed Research and Development Program of Oak Ridge National Laboratory, managed by UT-Battelle, LLC, for the U.S. Department of Energy.

\end{acknowledgments}

\bibliographystyle{apsrev4-2}
\bibliography{references}

\end{document}